\documentclass[twocolumn,showpacs,amsmath]{revtex4}

\usepackage{graphicx}%
\usepackage{dcolumn}
\usepackage{amsmath}

\begin{document}

\title{Interaction of vector solitons with a nonlinear interface}

\author{Ilya Shadrivov}

\affiliation{Nonlinear Physics Group, Research School of Physical
Sciences and Engineering, Australian National University,
Canberra ACT 0200, Australia} \email{ivs124@rsphysse.anu.edu.au}

\author{Alexander A. Zharov}

\affiliation{Institute for Physics of Microstructures, Russian
Academy of Sciences, Nizhny Novgorod 603950, Russia}
   \email{zharov@ipm.sci-nnov.ru}

\begin{abstract}
We develop the analytical method of field momenta for analyzing
the dynamics of optical vector solitons in photorefractive
nonlinear media. First, we derive the effective evolution
equations for the parameters of multi-component solitons composed
of incoherently coupled beams and investigate the soliton
internal oscillations associated with the relative motion of the soliton components. Then, we apply this
method for analyzing the vector soliton scattering by a nonlinear
interface. In particular, we show that a vector soliton can be
reflected, transmitted, captured, or split into separate components,
depending on the initial energy of its internal degree of freedom.
The results are verified by direct numerical simulations of
spatial optical solitons in photorefractive nonlinear media.
\end{abstract}
\maketitle

\section{Introduction}

Spatial solitons are self-trapped optical beams for which
diffraction divergence is exactly compensated by nonlinear
self-focusing \cite {pvs1}. Such solitons possess many important
properties which are believed to find a wide range of potential
applications in optical scanning, switching, and processing
devices. Spatial solitons can be generated in photorefractive
crystals at low powers (of order of $\mu W$) \cite{pvs3} and,
therefore, such materials look promising from a viewpoint of their use for the soliton generation in all-optical information circuits. 

One of the important properties of spatial solitons in photorefractive
media is their stability against collapse and break-up even in a
bulk medium due to the saturating nature of nonlinearity. Spatial
solitons in (1+1)-dimensional systems can be realized for the
light propagation in weakly guiding planar structures, where
self-focusing properties of the core material lead to the light
localization along the plane of a linear guiding structure.

Usually, a spatial soliton is viewed as a lowest-order
(fundamental) mode of an effective waveguide it excites in a
nonlinear medium. Moreover, higher-order modes of the induced
waveguide, that are associated with an effective
soliton-induced potential possessing more than one energy level, can be excited and such modes can be treated as the soliton internal modes.
Such internal modes are responsible for oscillations
of the soliton amplitude and width, and they have already been
studied in a number of theoretical papers \cite{pvs4, pvs5}.
A much more interesting situation takes place in the case of the
so-called vector spatial solitons, i.e. multi-component optical
solitons which consist of several mutually incoherent (or
orthogonally polarized) co-propagating beams. This problem got a little of attention in the past \cite{pvs20}, but it attracts much attention in view of recent experimental studies of multicomponent optical solitons.

The aim of the present paper is twofold. First, we study the
internal dynamics of (1+1)-dimensional vector solitons, associated
with the transverse relative oscillation of the centers of mass of
the partial soliton components. Second, for the first
time to our knowledge, we study the scattering of a vector soliton by a
nonlinear interface and demonstrate that the scattering process can be
modified dramatically by initial excitation of the soliton
internal modes.

As an example, we consider the dynamics of solitons in
photorefractive nonlinear media taking into account saturation of
the medium nonlinearity. A standard way for studying the excited
states of solitons is the analysis of the corresponding linear
eigenvalue problem for small perturbations. However, in the case
of nonintegrable systems, such excited eigenstates can be found
only by rather complicated computer simulations (see, for example,
Ref. \cite{pvs7} and references therein). In contrast, here we suggest
the method of momenta and applied it to analize the scattering problem. As
a matter of fact, such a method corresponds to the so-called
quasi-classical approximation in quantum mechanics. Regardless of
the fact that the method of momenta leads to the continuous
spectrum of the eigenstates, it allows us to consider large
perturbations (including the vector soliton splitting into individual 
components), since this approach is dealing with average
characteristics of a vector soliton.


\section{\label{pvss2}Vector spatial solitons.\protect\\ The method of momenta}

We consider spatial vector solitons consisting of $N$ incoherently coupled
components, which are described by the following set 
of Schr\"{o}dinger-like coupled
nonlinear equations (see. e.g. Ref. \cite{pvs8} and references therein):
\begin{equation}
\label{pvs1}
i\frac{\partial\Psi_j}{\partial x}+\frac{\partial^2\Psi_j}{\partial
z^2}-\lambda_j\Psi_j+\epsilon_{NL}\left(I\right)\Psi_j=0,
\end{equation}
where $x$ and $z$ are the propogation and transverse coordinates, respectively, $\Psi_j$ ($j =1,...N$) are the dimensionless slowly varying amplitudes of the beam components, $\lambda_j$ are the partial propagation
constants, $\epsilon_{NL}\left(I\right)$ is the nonlinear correction to the
dielectric permittivity, $I=\sum_{j=1}^N I_j$ is the total intensity of
the beam, and $I_j=\left|\Psi_j\right|^2$ are the partial intensities of the
soliton components. We assume that the nonlinear correction to the dielectric
permittivity depends only on the total intensity of the light, $I$, which can be written
approximately as follows \cite{pvs9}:
\begin{equation}
\label{pvs2}
\epsilon_{NL}(I)=I/\left(1+sI\right),
\end{equation}
where $s$ is the saturation parameter. A ground state of a vector spatial
soliton corresponds to the stationary solution of eq. (\ref{pvs1}) with
$\partial /\partial x = 0$ in which
$\Psi_j=\Psi_j^{(0)}\left(z-Z_j\right)$,
where $\Psi_j^{(0)}\left(z\right)$ are \emph{real} localized functions and $Z_j$ is
the center of mass of the $j$-th component. As a rule, the values
$Z_j$ for the vector solitons in the ground state are equal to each other and coincide with its center of mass. The dynamical behaviour of an excited vector soliton, such as oscillation of its amplitude and width, oscillation of the centers of mass of its components $Z_j$, or superposition of these oscillations, depends on initially excited intrinsic degrees of freedom. Here, it is important to note an analogy between a vector spatial soliton and a bound state of quantum particles. Indeed, the difference lies only in self-consistent nonlinear nature of the potential that provides the confinement for soliton components.

To study excited states of vector solitons we use the method of momenta that has been initially developed for scalar solitons of a Kerr-type nonlinear medium \cite{pvs10}. Here we generalize this aproach and technique to study multi-component spatial solitons in photorefractive media. 

The method
of momenta describes the electromagnetic beams in terms of integral field
characteristics, or field momenta \cite{pvs10}. The first momentum is defined as
\begin{equation}
\label{pvs3}
P_j=\int_{-\infty}^{\infty}I_j\,dz.
\end{equation}
It has the obvious physical meaning of the energy flux in the $j$-th soliton component. The second momentum is defined as
\begin{equation}
\label{pvs4}
\bar{Z_j}=\int_{-\infty}^{\infty}zI_j\,dz.
\end{equation}
Here the value $Z_j=\bar{Z_j}/P_j$ is the center of mass
of the $j$-th beam component. 

The purpose of this section is to derive a
close set of effective ordinary evolution equations for the momenta (\ref{pvs3}, \ref{pvs4}).
Similar procedure is used in semiclassical description of quantum systems (see, e.g. Ref. \cite{pvs11}). Let us multiply Eq. (\ref{pvs1}) by
$\Psi_j^*$ to obtain an equation for $P_j$, and by $z\Psi_j^*$ to obtain an equation for
$Z_j$, and then subtract the corresponding complex conjugated equations.
The integration of the resulting equations over $z$ yields
\begin{equation}
\label{pvs5}
\frac{dP_j}{dx}=0,
\end{equation}
\begin{equation}
\label{pvs6}
iP_j\frac{dZ_j}{dx}=\int_{-\infty}^{\infty}\left(\Psi_j^*\frac{\partial\Psi_j}{\partial z}-
\Psi_j\frac{\partial\Psi_j^*}{\partial z}\right)\,dz.
\end{equation}
Equation (\ref{pvs5}) means that the energy flux in each individual soliton component
is a conserved quantity. Thus, there is no energy exchange between the 
components. After differentiation of Eq. (\ref{pvs6}) by $x$ and subsequent
integration over $z$, we obtain
\begin{equation}
\label{pvs7}
P_j\frac{d^2Z_j}{dx^2}=\int_{-\infty}^{\infty}I_j\frac{\partial\epsilon_{NL}}{\partial z}dz.
\end{equation}
 The physical meaning of the Eqs. (\ref{pvs7}) comes from an analogy with the geometrical optics: the Eqs. (\ref{pvs7}) determine optical rays that correspond to the partial beams forming a vector soliton. From the set of equations (\ref{pvs7}) one can see that the center of mass of the whole beam moves along a straight line:
\begin{equation}
\label{pvs8}
\frac{d^2Z_c}{dx^2}=\frac{d^2}{dx^2}\left(\frac{Z_1P_1+Z_2P_2+\ldots+Z_NP_N}{P_1+P_2+\ldots+P_N}\right)=0.
\end{equation}

The set of Eqs. (\ref{pvs7}) is not closed. It cannot be used directly
since, in order to trace the dynamics of the soliton components, we should know the
electric field at each point $(x,z)$ in space, i.e. effectively complete
solutions of the Eqs. (\ref{pvs1}) are required. In this sense Eqs. (\ref{pvs7})
are equivalent to Eqs. (\ref{pvs1}), and so far we have not gained anything in addition. There are two ways to derive a closed set of equations from Eqs. (\ref{pvs7}). The first one is to obtain equations for higher order momenta which, nonetheless, does not guarantee that the system of equations will become closed.
A simpler way, which allows us to obtain the desired result, is to employ
a trial function to be substituted into Eqs. (\ref{pvs7}) instead of the unknown functions $\Psi_j$. It is clear that, for small deviations of the soliton constituents from the lowest-order (ground) equilibrium state, we can choose the trial functions in the form of unperturbed
stationary soliton components. Moreover, since within Eqs. (\ref{pvs7}) we
are operating with the average (integral) characteristics of the soliton, one can
guess that the result of the integration will not be sensitive to variations of the
transverse soliton structure, even for large deviations from the ground state.
This is a rather strong statement, and it will be verified below by means of direct
numerical simulations of the Eqs. (\ref{pvs1}).

\section{Two-component solitons}

To demonstrate the internal dynamics of the soliton excited states, we consider
a particular case of two-component vector solitons. As discussed above, it is natural
to choose the trial function for the Eqs. (\ref{pvs7}) in the form of a stationary solution $\Psi_j^{(0)}(z-Z_j)$. According to Eqs. (\ref{pvs7}), the relative shift between the soliton components,
$\Delta=Z_2-Z_1$, is governed by the following equation:
\begin{figure}
\includegraphics[width=8cm]{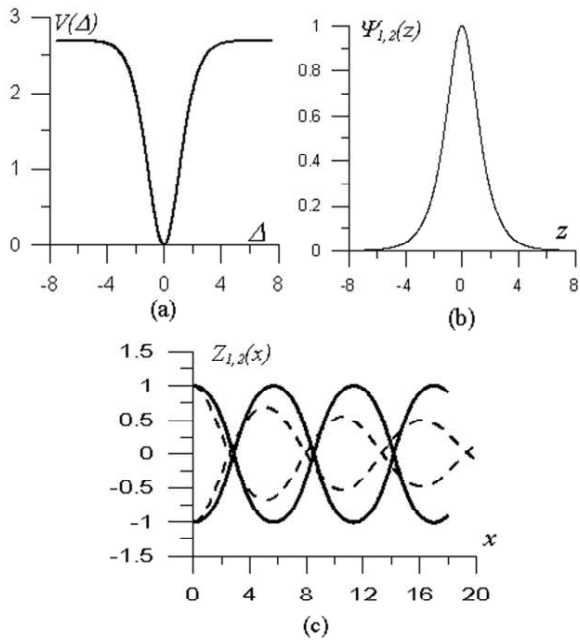}
\caption{\label{pvsf1}(a) Effective potential of the internal motion of the
soliton components, (b) transverse structure of the soliton with
identical components, (c) solid -- trajectories of the centers of mass of
the beams calculated by the method of momenta; dashed -- the same values found
by numerical simulations of the Eqs. (\ref{pvs1}).}
\end{figure}
\begin{figure}
\includegraphics[width=8cm]{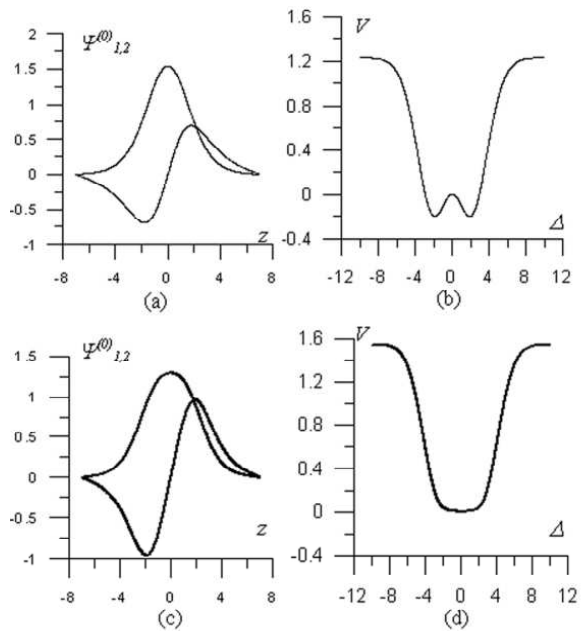}
\caption{\label{pvsf2}(a), (c) Structure of the soliton components at
$s=0.3$, $\lambda_1=1, \lambda_2=0.88$ and $\lambda_1=1, \lambda_2=0.9$, respectively. (b),
(d) Effective potentials of the internal motion for (a) and (c) solitons,
respectively.}
\end{figure}
\begin{figure}
\includegraphics[width=8cm]{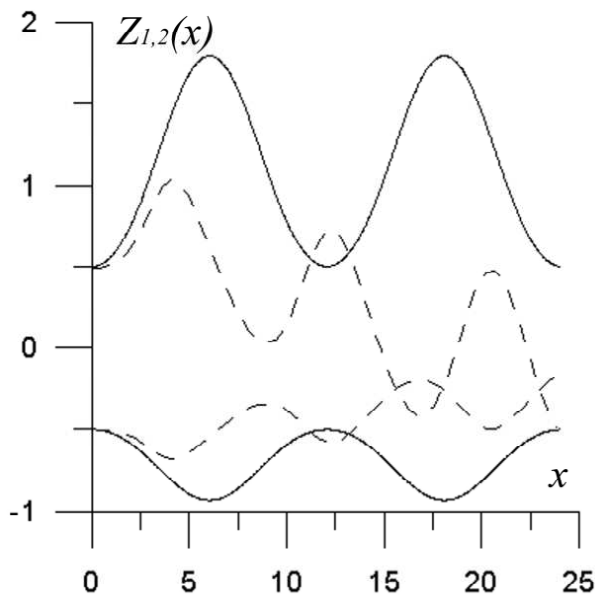}
\caption{\label{pvsf3} Dynamics of a two-hump soliton [shown in Fig. \ref{pvsf2} (a)] dynamics: solid -- trajectories calculated by the method of momenta; dashed
-- the same values found by numerical simulations of the Eqs. (\ref{pvs1}).}
\end{figure}
\begin{equation}
\label{pvs9}
\bar{P}\frac{d^2\Delta}{dx^2}=F\left(\Delta\right),
\end{equation}
 where $\bar{P}=P_1P_2/(P_1+P_2)$ is the reduced mass,
and
\begin{equation}
\label{pvs10}
F(\Delta)=2\int_{-\infty}^{\infty}I_2(z+\Delta)\frac{
\partial\epsilon_{NL}\left[I_1(z)+I_2(z+\Delta)\right]}{\partial z}dz.
\end{equation}
 Let us now introduce the effective potential of the beams interaction as follows 
\begin{equation}
\label{pvs11}
F(\Delta)=-\frac{dV(\Delta)}{d\Delta}.
\end{equation}
 We have now come to the simple mechanical model for the
internal dynamics of the two-component soliton, which describes two interacting quasi-particles in a bound state. This state can be thought of as a "photon molecule". Fig. \ref{pvsf1} shows the
effective potential $V(\Delta)$ for numerically calculated one-hump
soliton with two identical components. The minimum of the effective potential
corresponds to the stationary soliton in a ground state. The internal
oscillations of excited soliton components deducted from the solution of
Eq. (\ref{pvs9}) (shown in Fig. \ref{pvsf1} (c) by solid lines) are compared to the numerical solution of eqs.(\ref{pvs1}) (dashed lines in Fig. \ref{pvsf1} (c)).
The damping of the oscillations occurs due to radiation losses of the energy from the soliton originating from bends of the soliton component trajectories, which are not taken into account in the method of momenta.

 The next example deals with the two-hump solitons with the transverse structures of the components shown in Fig. \ref{pvsf2} (a,c) \cite{pvs12}. One can see that, when the amplitude of the two-hump component is significantly smaller than the amplitude of the
one-hump one, the stationary state with coaxial constituents is unstable because of the maximum of the potential at $\Delta=0$ (see Fig. \ref{pvsf2} (b)). It is interesting to note that, in this case, the ground equilibrium state of the
soliton corresponds to the soliton components with displaced centers. Such (1+1)D asymmetric vector solitons are well known from the theory of the two-component Kerr solitons described by Manakov model ($s\to 0$ in eqn. \ref{pvs2}) \cite{pvs14} and for the partially coherent multi-component solitons in photorefractive medium \cite{pvs15}. However, when the amplitudes of the components are comparable, the coaxial state becomes stable again (see Fig. \ref{pvsf2} (d)).
The dynamics of the unstable coaxial soliton state calculated by solving eqs.(\ref{pvs9}), as well as eqs.(\ref{pvs1}), is shown in Fig. \ref{pvsf3}.
One can see that, in this case, the agreement between the method of momenta and numerical simulations is much worse than for the one-hump solitons because of the strong reshaping of the two-hump beams. This reshaping, along with the radiation processes, violates the assumptions of the analytical method.

\section{Interaction of vector solitons with a nonlinear
interface}

The method of momenta looks promising for treating the problem of vector soliton
dynamics in inhomogeneous media. In this section we consider interaction of a two-component
vector soliton with an interface between two photorefractive media. Both
linear and nonlinear properties of the materials are discontinuous on the interface. The problem of one-component photorefractive soliton interaction with nonlinear interface had
recently been studied in \cite{pvs13}, soliton interaction with interface in the Kerr-like medium can be found in \cite{pvs16, pvs17}. Let us represent the coordinate
dependences of the partial propagation constants and nonlinear correction to dielectric
permittivity by the following form:
\begin{eqnarray}
\label{pvs12}
\lambda_{1,2}(z)=\lambda_{1,2}+\delta\lambda_{1,2}\cdot 1(z),\nonumber\\
\epsilon_{NL}(z)=\epsilon_{NL}(I)\cdot\left[1+\delta\epsilon_{NL}\cdot
1(z)\right]
\end{eqnarray}
where $1(z)=\left\{
 \begin{array}{rl} 1, \mbox{} z>0 \\0, \mbox{} z<0
\end{array} \right.$
is the unit step function. By the method of momenta one can obtain that the
energy fluxes in both
beams are conserved.
For the centers of mass of the components we have:
\begin{eqnarray}
\label{pvs13}
P_{1,2}\frac{d^2Z_{1,2}}{dx^2}=\int_{-\infty}^{\infty}\left[2+\delta\epsilon_{NL}\cdot1(z)
\right]I_{1,2}\frac{\partial\epsilon_{NL}(I)}{\partial
z}\,dz-\nonumber\\
-\delta\epsilon_{NL}I_{1,2}(-Z_{1,2})+\delta\lambda_{1,2}I_{1,2}(-Z_{1,2})\times\nonumber\\
\times\epsilon_{NL}\left[I_{1,2}(-Z_{1,2}) +I_{2,1}(-Z_{2,1})\right]\nonumber\\
\end{eqnarray}
All the terms in the Eqs. (\ref{pvs13}) have clear physical meaning. The terms
that do not contain the jumps of the medium parameters represent the internal forces
acting between soliton components. They are responsible for an internal
dynamics of the soliton and, in the homogeneous case, can be reduced to
the equation (\ref{pvs9}). The terms proportional to
$\delta\lambda_{1,2}$ and $\delta\epsilon_{NL}$ are the
external forces perturbing the internal oscillations of the soliton components
at the collisions of quasi-particles with an interface. 

To derive the closed set
of equations describing the interaction of a vector soliton with an interface we substitute into Eqs. (\ref{pvs13}) trial functions
in the form of the stationary unperturbed soliton component (see Section \ref{pvss2}). One can expect that this ansatz will be sufficiently accurate whenever transverse structures
of the beams are not strongly affected by large discontinuities of the
medium parameters. We will verify this assumption by comparing direct computer
simulation of the wave equations (\ref{pvs1}) (in the case of
inhomogeneous media) with the results of quasi-particle theory obtained
with the method of momenta. In what follows, we study the following
effects of soliton interaction with an interface: (a)
excitation of internal degrees of freedom; (b) the change of the type of soliton interaction with an interface (i.e. reflection, transmission and capture) as function of
the initial internal energy of the vector soliton launched under different
initial conditions; (c) the splitting of pre-excited soliton into individual
components at the collision with an interface. It should be noted, with regard to the point (a) above, that a spatial soliton consisting of identical components, being
initially in ground state, does not become excited at the collision with an interface. This happens because, due to identical character of interaction of every
component with interface, the coaxial structure of the soliton state is preserved. 

\begin{figure}
\includegraphics[width=8cm]{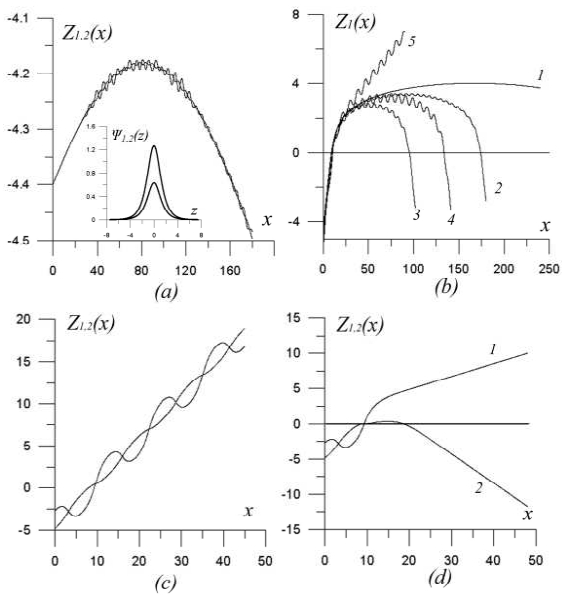}
\caption{\label{pvsf4} (a) Excitation of the internal oscillations of the soliton
at the reflection from the interface, inset shows the transverse structure of the soliton, (b) soliton interaction with the interface
at different values of $\Delta$: 1 -- $\Delta=0.0$, 2 -- $\Delta=0.4$, 3 -- $\Delta=0.8$, 4 -- $\Delta=1.2$, 5 -- $\Delta=1.4$, (c) $\Delta=2.2$, beams make up a bound state without interface, (d) beams are split at the interface.}
\end{figure}

The results that we present in this section are obtained
for the one-hump soliton shown in the inset in Fig. \ref{pvsf4} (a) and interface parameters are as
follows: $\delta\lambda_{1,2}=-0.15$; $\delta\epsilon_{NL}=0.2$; $ds=0$.
\begin{figure}
\includegraphics[width=8cm]{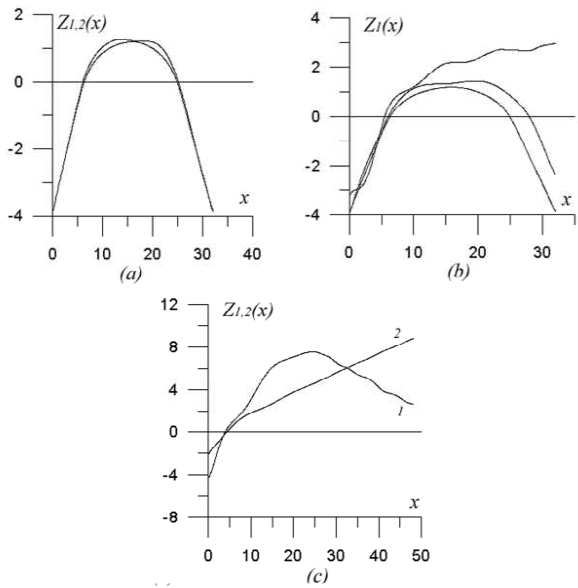}
\caption{\label{pvsf5}Numerical simulations: (a) excitation of the internal oscillations of the soliton
at the reflection from the interface, (b) soliton interaction with the interface
at different values of initial value of $\Delta$, (c) separation of the beams at the interface, $\Delta=2.2$.}
\end{figure}
The solutions of the ordinary differential
equations (\ref{pvs13}) are shown in Fig. \ref{pvsf4}.
 Excitation of internal oscillations after soliton reflection from an
interface is demonstrated in Fig. \ref{pvsf4} (a). Soliton dynamics
near the interface depends dramatically  on the initial energy
level of the internal oscillations of the components. In
Fig. \ref{pvsf4} (b) soliton components were launched to the
interface at the same angle $\left( dZ_{1,2}/dx=-0.5\right)$ and
from the same position of the common center of mass. The only
adjustable parameter, which determines the initial internal energy
of oscillations, was the relative shift between components
$\Delta$. One can see that, depending on the value of $\Delta$,
the transmission (curve 5 -- $\Delta=0.12$), reflection (curves
2,3,4 -- $\Delta=0.4, 0.8, 1.2$, respectively) and capture into the
unstable nonlinear surface mode (curve 5 -- $\Delta = 0.0$) take
place. 

\begin{figure}
\includegraphics[width=8cm]{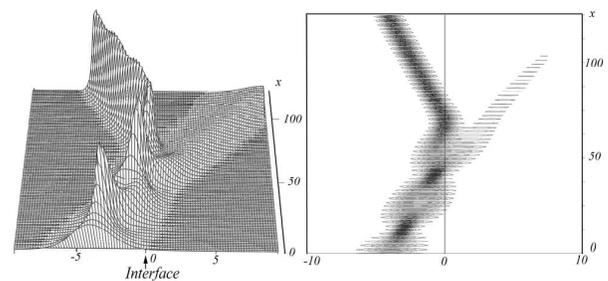}
\caption{\label{pvsf6} Numerical simulations: strong reshaping of the two-hump soliton at the interaction with an interface takes place, surface and contour plots. Initial conditions are taken in the form of the soliton shown in Fig. \ref{pvsf2} (a). Parameters of the interface are the same as in Fig. \ref{pvsf5}.}
\end{figure}
It is interesting to note that the character of soliton
interaction depends upon the parameter $\Delta$ non-monotonically, which
can be explained by considering the phase of the
internal oscillations at the interface. This phase depends on the
initial value of $\Delta$ by virtue of anharmonicity of internal
oscillations in the effective potential well. Fig. \ref{pvsf5}
demonstrates the results of numerical calculations of equations
(\ref{pvs1}), in which $\lambda_{1,2}$ and $\epsilon_{NL}$ are
taken according to eq. (12). Excitation of the internal degree of
freedom takes place at the reflection from the interface
Fig. \ref{pvsf5} (a). The calculations shown in Fig. \ref{pvsf5} (b)
were carried out for the same initial conditions as for
curves 1,2, and 3 in Fig. \ref{pvsf4} (b). The higher values of the
initial relative shift $\Delta$ between individual components do not result in the same
behaviour of the trajectories as in Fig. \ref{pvsf4} (b). It can be
easily explained by the strong damping of oscillations due to the
radiation effects, which takes place for large values $\Delta$.
 As a result, the amplitude of
the internal oscillations is small at the point where soliton
meets the interface and the outcome of interaction is very
similar to the one for the smaller initial values of $\Delta$. 
To conclude, one can check that method of momenta leads to the satisfactory
agreement with the direct numerical solution of nonlinear Schr\"{o}dinger
equations for the values of $\Delta$ up to the half width of the
soliton. 

The splitting of the vector soliton into individual components is
shown in Fig. \ref{pvsf4}. We have revealed that the splitting
occurs at the angles of incidence of the soliton beam close to the angle
of total internal reflection and for large enough values of initial
relative shift between components ($\Delta=2.2$,
$dZ_1/dx=dZ_2/dx=-0.5$). Without interface, the soliton components
make up a bound state (Fig. \ref{pvsf4} (c)). At the presence of
interface the method of momenta predicts that splitting into components takes place
(Fig. \ref{pvsf4} (d)). This effect can be treated in terms of the
quasi-particle theory. Indeed, when the depth of effective potential
well is of the same order that the energy level of the bound
state, than even a small perturbation (at the interface) of this state can lead to its
destruction. 

We have obtained the similar splitting effect
[shown in Fig. \ref{pvsf5} (c)] by solving the equations (\ref{pvs1}) numerically for 
the initial conditions used in the method of momenta. In numerical simulations, we can observe the separation of the beam components, but in the case of such large
values of the initial transverse shift, the value of the first
momentum (\ref{pvs4}) fails to determine the position of the center
of mass of the localized beams. Significant part of the energy is
transferred into radiation, and the integration in
(\ref{pvs4}) gives us the coordinate of the common center of mass
of both localized and nonlocalized waves. It is clear that the
less the radiation losses are, the better the position of the localized beam
is described by (\ref{pvs4}).

Finally, it is neccessary to note that the dramatic reshaping of
the multi-hump solitons at the interaction with the interface
takes place. Numerical simulations show that these complex beams
can be transformed into a set of single hump solitons and a
considerable part of the electromagnetic energy goes into the
radiation even for a very small change of the parameters of the
medium at the interface (see Fig. \ref{pvsf6}). This reshaping makes it impossible to apply the method of momenta to the problems of multi-hump soliton dynamics in the
inhomogeneous media.


\section{Conclusions}

We have developed the method of momenta to analyze the dynamics
of vector solitons consisting
of two (or more) incoherently coupled components, in both homogeneous and inhomogeneous photorefractive nonlinear media. First, we have
demonstrated that this method provides an effective
analytical tool for the study of internal oscillations of 
vector solitons. Second, we have studied the scattering of a
vector soliton by a nonlinear interface, and we have demonstrated
that this type of the soliton interaction can lead to the
excitation of the soliton's internal oscillations. In particular,
we have demonstrated that the same vector soliton incident onto
an interface can become reflected, transmitted, captured by the
interface, or even split into its constituents, depending solely on
the energy of the initially excited internal oscillations.

\begin{acknowledgments}
Authors wish to thank Yu.S.Kivshar and E.A.Ostrovskaya for the help in the preparation of the article.
\end{acknowledgments}
\bibliography{interaction}
\end{document}